# Detection and modeling of carrier capture by single point defects under variable electric fields

Artur Lozovoi[1], YunHeng Chen[3], Gyorgy Vizkelethy[4], Edward Bielejec[4], Johannes Flick[1,5], Marcus W. Doherty[3], Carlos A. Meriles[1,2,†]

[1]*Department. of Physics, CUNY-City College of New York, New York, NY 10031, USA.* [2]*CUNY-Graduate Center, New York, NY 10016, USA.* [3]*Department of Quantum Science and Technology, Research School of Physics, Australian National University, Canberra, Australian Capital Territory 2601, Australia.* [4]*Sandia National Laboratories, Albuquerque, New Mexico 87185, USA* [5]*Center for Computational Quantum Physics, Flatiron Institute, New York, NY 10010, USA.*

[†]*Corresponding author. E-mail:* cmeriles@ccny.cuny.edu.

**ABSTRACT**: Understanding carrier trapping in solids has proven key to the development of semiconductor technologies but observations thus far have relied on ensembles of point defects, where the impact of neighboring traps or carrier screening is often important. Here, we investigate the capture of photo-generated holes by an individual negatively-charged nitrogen-vacancy (NV) center in diamond at room temperature. Using an externally gated potential to minimize space-charge effects, we measure the capture probability under electric fields of variable sign and amplitude to find an asymmetric-bell-shaped response with maximum at zero voltage. To interpret these observations, we run semi-classical Monte Carlo simulations modeling carrier trapping through a cascade process of phonon emission, and obtain electric-field-dependent capture probabilities in good agreement with experiment. Since the mechanisms at play are insensitive to the characteristics of the trap, we anticipate the capture cross sections we observe — largely exceeding those derived from ensemble measurements — may also be present in materials platforms other than diamond.

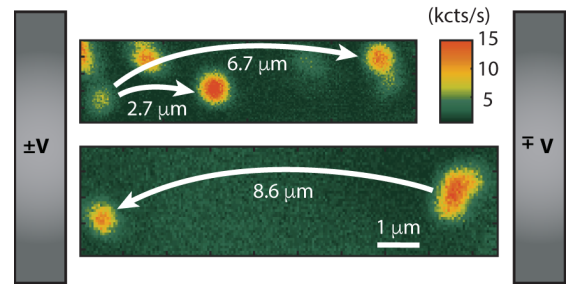

Given the important role of carrier capture on the functionality of semiconductor devices, the last decades have seen much effort devoted to developing experimental and computational tools able to shed light on the physical mechanisms governing charge trapping dynamics[1-3]. This interest has broadened in recent years owing to the promise of semiconductor-hosted point defects for applications in nanoscale sensing and quantum information processing[4,5], as electric field fluctuations from photo-induced ionization and carrier recapture are, e.g., responsible for the instabilities observed in the emission spectra of single-photon sources[6-8] Along similar lines, carrier trapping can negatively impact spin qubit detection schemes based on the monitoring of photo-generated carriers[9-12]. An in-depth understanding of charge transport and capture is also central to the realization of quantum information processing protocols where spin-polarized carriers serve as a bus to communicate remote solid-state qubits[13].

Traditionally, capture cross sections have been extracted from mesoscale transport measurements in crystals with relatively high concentrations of point defects, where unintended effects (emerging, e.g., from proximity between charge traps or capture by point defects other than those targeted) are difficult to account for. Deriving capture cross sections is especially challenging for insulators due to the lack of a drain path for electrical current. Note that while the use of known dopants can mitigate the problem, screening effects can be severe when the donor and/or acceptor content is high to moderate, which, e.g., has a particularly deleterious impact on the investigation of carrier capture by hydrogenic point defects.

Many of these limitations, however, can be circumvented with the help of nanoscale probes sensitive to local changes in the concentration of trapped charge. For example, scanning tunneling microscopy was recently exploited to identify and characterize carrier trapping by individual point defects in hexagonal boron nitride upon the application of voltage pulses[14], an ability subsequently extended to demonstrate writing, reading, and erasing of doping patterns with nanometer resolution[15]. By the same token, confocal microscopy experiments on individual negatively charged nitrogen-vacancy (NV⁻) centers in diamond reported the capture of itinerant carriers photo-injected microns away, a result attributed to the formation of transient Rydberg states[16]. A similar technique was recently applied to investigate the charge dynamics of silicon-vacancy (SiV) centers in diamond under optical excitation[17,18].

In this work, we experimentally examine hole transport between individually addressable NV centers in a pristine diamond crystal under externally applied electric fields.

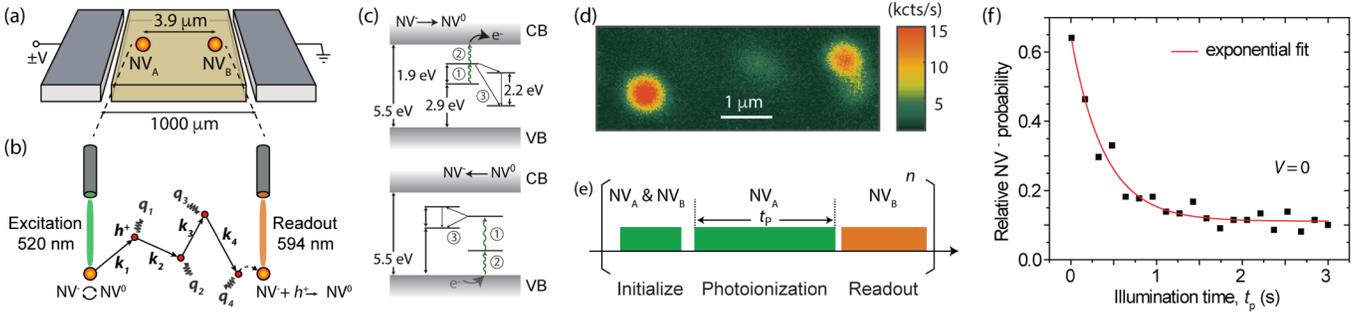

**Figure 1. Carrier transport between single NV centers.** (a) We investigate photo-activated carrier transport between two individual NVs in the presence of an external electric field. (b) Illustration of hole diffusion mediated by phonon scattering. $k_i$ and $q_i$ stand for the hole and phonon wavevectors, respectively. (c) Energy level diagrams of NV$^-$ and NV$^0$ describing the two-photon ionization and recombination processes. Continuous green illumination leads to the creation of free electrons and holes (upper and lower panels, respectively). Numbers in circles indicate stages in each process; VB and CB denote the valence and conduction bands. (d) Confocal scan of NV$_A$ and NV$_B$ under 0.5 mW, 520 nm illumination. (e) Pulse sequence. (f) NV$^-$ population as a function of the photoionization pulse duration, $t_p$ at 1mW. The power and duration of the initialization (readout) pulse are respectively 1 mW and 100 μs (10 μW and 20 ms). All experiments are carried out at room temperature.

Using one of the color centers as a point of carrier photo-injection and the other as a negatively charged trap, we find that the presence of a bias field quickly reduces the hole capture probability, even in the case where the drift force pushes the carrier towards the trap. We numerically reproduce these observations via a Monte Carlo simulation simultaneously encapsulating hole diffusion, drift, and capture.

**Results.** *Photo-activated charge transport between individual color centers.* We use a focused ion beam to implant an electronic grade diamond (2×2×0.1 mm, Diamond Delaware Knives) with 20 MeV N$^+$, an energy sufficient to propel the nitrogen ions ~10 μm into the crystal. Subsequent high-temperature annealing[19] resulted in the formation of spatially-resolved NV islands of varying concentrations, down to single color centers. For the purposes of this work, we first focus on a pair of NVs 3.9 μm apart (Figs. 1a through 1c), which we use to investigate carrier capture following a protocol similar to that implemented before[16]. Briefly, we use a 520-nm laser beam to preferentially initialize both NVs into the negatively charged state (approximately with 75% chance). Subsequent green illumination of one of the NVs in the pair — here referred to as NV$_A$ or the "source" NV — produces multiple instances of NV interconversion between the negative and neutral charge states[20]. Since every cycle of NV ionization and recombination leads to the successive injection of a free electron and a hole (see level diagram in Fig. 1d), this stage in the protocol results in a stream of diffusing carriers, which we exploit to alter the charge state of the twin color center in the pair, namely, NV$_B$ or the "target" NV. To probe its charge state after a variable exposure time, we implement a single-shot charge state readout with the help of a low-power, 594-nm laser[21,22] (see Fig. S1 of the Supplemental Material); all experiments are carried out at room temperature.

Unlike the case of a hole approaching an NV$^-$, no Coulomb interaction acts on an electron proximal to a neutral NV. Correspondingly, the charge conversion of NV$_B$ is effectively a one-directional process, from NV$^-$ to NV$^0$, despite the balanced numbers of photo-injected carriers with opposite signs. An example experimental curve obtained using this approach is shown in Fig. 1f where we plot the relative fluorescence of the target as a function of the photoionization laser pulse duration, $t_P$. The observed decay reveals the increasing probability for NV$_B$ to capture one of the diffusing holes photo-generated out of NV$_A$ and thereby change its charge state into neutral (which does not fluoresce under 594 nm illumination). In line with this interpretation, the bleaching rate extracted from the exponential fit can be thought of as the unit-time hole capture probability, and thus used to calculate the hole capture cross section in the assumption of isotropic diffusion out of NV$_A$[16]. The obtained value $\sigma_h^{(Exp)} \sim 3 \times 10^{-11}$ cm$^2$ is 2 to 4 orders of magnitude greater than those reported before for Coulombic attractive traps[27,28], likely the combined consequence of the high sample purity (and hence large inter-defect separation), the ability to probe individual point defects (as opposed to ensembles), and the large bandgap of diamond (leading to unscreened interactions between the charge carrier and the trap).

In order to gain a better understanding of the hole transport and capture process, we extend these prior experiments to investigate the system response in the presence of an externally applied electric field. Fig. 2a shows our protocol: We use parallel surface electrodes to apply a constant voltage of variable amplitude and sign during the photoionization pulse (i.e., during carrier transport). The geometry of the electrodes makes the electric field lines roughly aligned with the line connecting NV$_A$ and NV$_B$ (Fig. 1a). The upper plot in Fig. 2b displays the measured bleaching rates — or, equivalently, the hole capture probability — as a function of the applied electric field, with positive values corresponding to the direction of hole acceleration from $NV_A$ towards $NV_B$. The response peaks at 0 V·m$^{-1}$ to then decay slightly asymmetrically with increasing voltage in both the positive and negative directions.

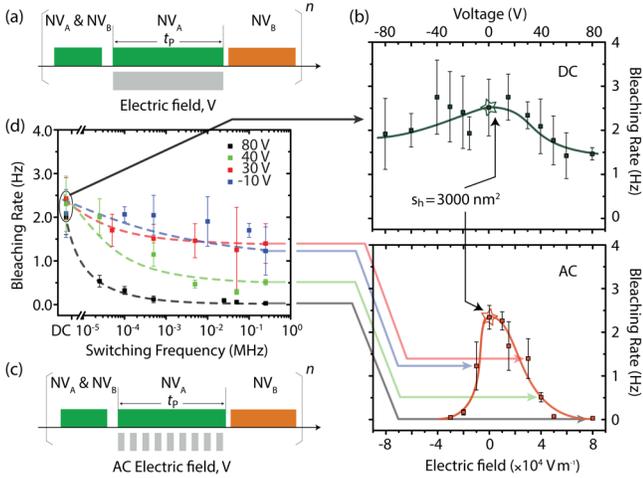

**Figure 2. Hole capture under DC and AC electric fields.** (a) Schematic of the DC protocol. (b) Hole capture unit-time probability as a function of the applied DC voltage (top plot, dark green squares) or as extracted from the plateau values in the AC protocol (lower plot, orange squares). Solid lines are guides to the eye. (c) Same as in (a) but for an AC electric field with 50% duty cycle. (d) Bleaching rate of the target NV under an AC electric field of +80 V, +40 V, +30 V and -10 V at variable frequency; solid horizontal segments indicate the plateau values in the high AC frequency limit. First points (labeled "DC") correspond to the experiments with the voltage constantly on.

Prior work has shown[29] that the presence of a constant (DC) electric field typically leads to the formation of metastable space charge potentials that can be locally comparable in strength to the external field (and thus can effectively shield it). In the current set of experiments, this results in the observation of distorted bleaching rates under the applied voltage, often dependent on the initialization protocol and measurement history. To mitigate this complication, we conduct a series of measurements where the electric field is switched on and off at various frequencies with a 50% duty cycle during the photoionization laser pulse (Fig. 2c and lower plot in Fig. 2b). This alternating current (AC) protocol strongly suppresses the formation of space charge potentials[29], hence allowing us to determine the capture probability more faithfully. Fig. 2d shows the bleaching rate as a function of the switching frequency for different inter-electrode voltages. We find a gradual decrease with increasing switching frequency to reach voltage-dependent plateaus at frequencies greater than ~10 kHz. Comparing these limit AC bleaching rates against those derived from the DC experiments, we observe an overall reduction at non-zero voltages, even in the case where the applied electric field accelerates holes towards the target NV (corresponding to positive voltages).

To rationalize this behavior, we first note that the space charge potentials formed by carriers captured during periods of photo-injection with the electric field "on" get quickly neutralized when the voltage is "off". Therefore, increasing the on/off frequency progressively shortens the time interval available for space charge formation, hence facilitating the propagation of all photo-injected carriers. The fact that the bleaching rate stops changing above ~10 kHz supports this model and tells us that the formation of the space charge distribution at a given voltage takes longer than ~100 µs. In line with these ideas, we interpret the limit value at higher switching frequencies as the space-charge-compensated bleaching rate, which we hereafter take as the most representative unit-time hole capture probability. We caution that the values we derive must still be seen as an upper bound because bleaching of the target NV necessarily takes place during the voltage-free intervals, thus making the capture rates slightly greater than otherwise.

It is worth noting that the presence of AC electric fields has been seen to alter the ionization rate of the di-vacancy and silicon-vacancy centers in SiC[30]. This effect can potentially interfere with the interpretation of our experimental data as our approach assumes voltage-insensitive carrier injection rates. We carried out NV recombination and ionization rate measurements in the presence of DC and AC electric fields in the range of amplitudes and frequencies corresponding to the experiments in Fig. 2, and observed no change[22] (see Fig. S5 in the Supplemental Material). We attribute this result to the high purity of the sample we used: Indeed, the ionization rate change observed in Ref. [30] was assigned to the charge state modulation of the nearby defects in the presence of photo-ionized carriers and electric field.

*Carrier drift at variable distance.* To test the universality of the behavior shown in Fig. 2b, we extend the transport experiments under AC electric fields to three more pairs of NVs separated by different distances (Fig. 3a). Fig. 3b shows the corresponding dependencies of the bleaching rate as a function of the AC field amplitude. All four pairs of NVs display similar behavior despite the difference in distance, location, and direction of carrier propagation relative to the electric field orientation.

Fig. 3c shows the dependence of the bleaching rate in the absence of electric field, i.e. $V = 0$, as a function of distance $d$ between the NVs. It follows an inverse square dependence, which is consistent with the geometric argument of capture cross section over the sphere surface area scaling as a function of the sphere's radius. We note that is observation is complementary to the one reported recently in the same sample[16] as it adds new NV pairs thereby strengthening the case. We use the observed dependence to scale the data in Fig. 3b to the values measured at 3.9 µm (Fig. 2b). The main plot in Fig. 3b clearly demonstrates the universal character of the observed behavior: All curves peak at $V = 0$ and show a similar dependence with distance. Here, we note that the observed asymmetry of the curves with a longer tail at positive voltages is more pronounced for some of the pairs, which, in turn, was found to be affected by the geometry of the electrodes and the sample holder (see discussion in the Supplemental Material).

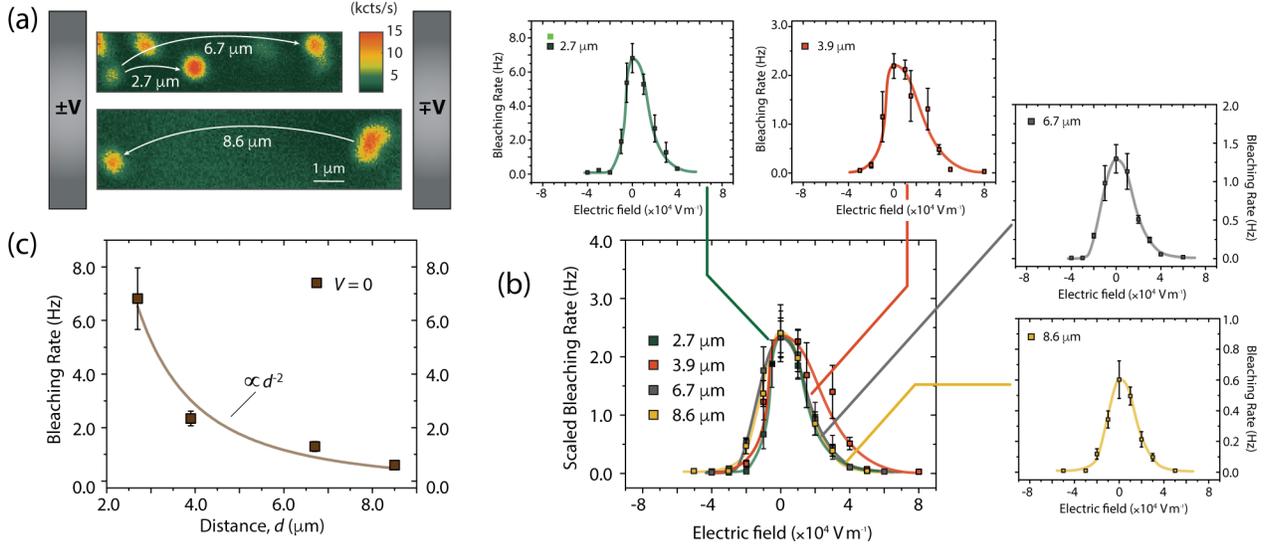

**Figure 3. Hole capture under AC protocol for different NV pairs.** (a) Confocal images of several NV pairs; arrows label the physical separation and show the direction from the source to the target NV. (b) Hole capture unit-time probability as a function of the applied AC voltage for four NV pairs. The main plot shows the same curves scaled to the values corresponding to the distance of 3.9 μm using an inverse square dependence. Positive voltages correspond to hole acceleration; solid lines are guides to the eye. (c) Hole capture unit-time probability as a function of the source/target distance for $V = 0$. The solid line is an inverse square fit.

*Monte Carlo simulations.* Developing a microscopic framework that simultaneously takes into account the mesoscale nature of the carrier transport and capture processes in our present experiments is a complex task. Carrier capture by deep defects has long been known to involve multi-phonon emission processes[31-36], but a first-principles description — valid for thermal equilibrium or under the action of an external field — is still the subject of much work[37-40]. The problem is especially hard for charged point defects — and more so in the absence of screening — because the range of distances required to accurately describe Coulombic interactions largely exceeds the supercell size typical in ab-initio methods. In the present case, carrier propagation over mesoscale distances adds to the problem's complexity because developing analytical solutions to the governing Boltzmann transport equation — with or without linearization of the external drift term — presents a formidable mathematical challenge.

The long range of the interactions at play, however, can also be seen as an advantage in the sense that the capture process becomes less sensitive to the specifics of the attracting trap. Further, the quasi-continuum in the hydrogenic series describing the energies of the trapped carrier at early stages allows for a semi-classical treatment of the capture process, as first discussed by Lax[31]. Within this framework, carrier capture takes place through a "cascade" emission involving acoustic and optical phonons (with the latter playing a gradually more important role at higher temperatures). Carrier capture takes place once the particle's energy falls below a "cutoff", typically chosen to be commensurate with the characteristic energy of optical phonons in the host crystal.

Building on these ideas, here we model our observations through a hybrid approach leveraging the Monte Carlo framework[41-44], already applied successfully to describe hole and electron transport in diamond[45-48]. Following diffusive propagation over a variable distance (see below), we incorporate the dynamics of hole capture by tracking the particle's energy, $E_0 = E_k + U$, as it undergoes inelastic phonon scattering. In the above expression, the two terms respectively denote the kinetic energy, $E_k = \hbar^2 k^2/2$, and the Coulomb interaction, $U = -e^2/(4\pi\varepsilon_0\epsilon r)$; $\boldsymbol{k}$ is the hole wavevector, $\varepsilon_0$ is the vacuum permittivity, $\epsilon = 5.4$ is the relative dielectric constant of diamond, $r$ is the (time-dependent) hole–trap distance, $e$ denotes the elementary charge, and $\hbar$ is the reduced Planck constant.

As illustrated in the energy diagram of Fig. 4a, the negative energy levels below the valence band correspond to bound hole states and can be crudely approximated by a hydrogenic series approaching the continuum at longer distances[16,49,50]. During a Monte Carlo realization, $E_0$ drops when the hole loses its kinetic energy via emitted phonons in the spatial region of the potential well created by the trap. In our simulations, we set the cutoff energy $E_c$ at 175 meV (red dashed line in Fig. 4a), which corresponds to the energy of an optical phonon in diamond[22,51].

Prior Monte Carlo simulations in Si and $SiO_2$ modeled carrier propagation and capture by point traps via Boltzmann transport equations under the assumption of high free carrier concentration[52-54]. This regime, however, is not directly applicable to the present work because the diamond sample we use contains no thermal carriers, even at room temperature. Further, since the NV charge state interconversion time (of order ~1 μs under our present conditions) is much longer than the hole diffusion time (~10 ns for a ~4 μm distance[55]), there can only be one free hole present in the vicinity of the source and target NVs at any given time. Correspondingly, we can employ Newtonian equations of motion in three dimensions to model hole transport with inelastic phonon scattering events introduced into the simulation statistically.

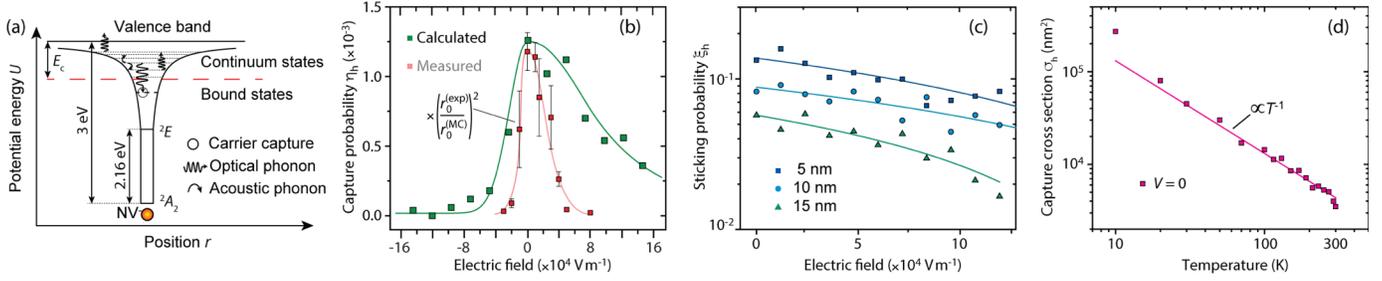

**Figure 4. Monte Carlo simulation results.** (a) Energy level diagram for a hole in the vicinity of an NV⁻ trap. We separate the hydrogenic series of states into a high-energy continuum and a discrete set based on a cutoff at $E_c = 7k_BT$ corresponding to the optical phonon energy in diamond (red dashed line). Wavy and curved arrows indicate possible phonon-mediated transitions between these levels. (b) Calculated hole capture probability as a function of the applied voltage (green squares). For comparison, the faint red squares show the experimental values as determined from Fig. 2 and the total number of photo-emitted holes. Given the longer inter-defect distance in the experiment compared to that in the simulation ($r_0^{(Exp)} = 3.9$ μm and $r_0^{(MC)} = 0.35$ μm, respectively), all measured values have been renormalized by the square of the ratio between distances. (c) Calculated fraction of captured holes $\xi_h$ from a set starting within a sphere of 5, 10, and 15 nm radius around the target. The sticking probability drops as the voltage increases due to the energy gain. (d) Temperature dependence of the unbiased hole capture cross section as calculated by Monte Carlo simulations. In (b) through (d), solid lines are guides to the eye.

Each Monte Carlo realization starts with a thermal hole moving in a random direction. Since simulation of transport over micrometer distances requires computational resources beyond state-of-the-art capabilities, we set a smaller initial distance (of up to 500 nm at room temperature) between the trap and the hole. In Fig. S4 of the Supplemental Material[22], we show simulation results for different initial starting points, and demonstrate that they converge to the same outcome at distances greater than 350 nm. Based on this finding, we quantify the hole capture probability as $\eta_h = N_c/N$, i.e., the ratio between the number of times $N_c$ that the hole gets captured by the target and the total number of realizations $N$. Fig. 4b shows $\eta_h$ (green dots) as a function of the externally applied electric field amplitude. Comparison with the measured capture probabilities — derived from the known hole emission rates[22] and the experimental data set in Fig. 2, faint red squares in Fig. 4b — shows a good correspondence both in overall shape and value. For an initial hole–trap distance $r_0 = 0.45$ μm, we find $\sigma_h^{(MC)} = 4\pi r_0^2 \eta_h(V=0) \approx 3.2 \times 10^{-11}$ cm² in good agreement with experiment (see also Fig. S4 of the Supplemental Material).

Since the number of carriers reaching NV$_B$ grows for positive voltages, the decay of the capture probability with external bias underscores the impact of the field on the capturing dynamics. We can more clearly see this interplay through the "sticking probability"[31] determined as the fraction of capture events from the total number of realizations where the hole reaches a small sphere of radius $d$ centered at the target NV (Fig. 4c). We find that even when $d \lesssim \sqrt{\sigma_h}$, a moderate electric field can have a substantial effect in reducing hole capture.

Lastly, we leverage our Monte Carlo model to calculate the (unbiased) hole capture cross section by NV⁻ at variable temperatures, and find $\sigma_h^{(Cal)}(T) \propto T^{-1}$ above 20 K (Fig. 4d). Physically, this response stems from the stable formation of larger bound exciton orbits at lower temperatures[16], and qualitatively reproduces the trend found in prior observations[56] (we note that the exponent governing the thermal dependence is different, a feature possibly related to the lower charge screening in our sample).

Combined with these large capture cross sections, the long mean free paths characteristic of high-purity, low-temperature diamond[57] promise intriguing opportunities to boost the hole trapping probability, e.g., through the implementation of time-resolved, synchronous protocols designed to decelerate the charge carrier as it reaches critical proximity to the target. By the same token, these results could serve as a stepping stone in addressing intriguing fundamental questions such as the ability to inject and probe the lifetime of spin polarized carriers, and the stability of Rydberg-like states forming upon hole capture. Besides the NV center, our experiments can be generalized to investigate trapping by other, non-fluorescent defects of importance. One example is the P1 center — a paramagnetic defect formed by substitutional nitrogen — whose charge state could be monitored using an adjacent NV as a local spin probe.

## ASSOCIATED CONTENT

**Supporting Information:** Contains additional information on sample characteristics and instrumentation used. It also provides additional details on the Monte Carlo modelling.

**AUTHOR INFORMATION**

**Corresponding author:**
[†]E-mail: cmeriles@ccny.cuny.edu
**Notes**
The authors declare no competing financial interests.

**ACKNOWLEDGMENTS**

A.L. and C.A.M acknowledge support from the National Science Foundation through grants NSF-1914945 and NSF-2216838; they also acknowledge access to the facilities and research infrastructure of the NSF CREST IDEALS, grant number NSF-HRD-1547830. M.W.D. acknowledges support from the Australian Research Council DE170100169. Ion implantation work to generate the NV and SiV centers was performed, in part, at the Center for Integrated Nanotechnologies, an Office of Science User Facility operated for the U.S. Department of Energy (DOE) Office of Science. Sandia National Laboratories is a multi-mission laboratory managed and operated by National Technology and Engineering Solutions of Sandia LLC, a wholly owned subsidiary of Honeywell International Inc., for the U.S. Department of Energy's National Nuclear Security Administration under contract DE-NA0003525. This paper describes objective technical results and analysis. Any subjective views or opinions that might be expressed in the paper do not necessarily represent the views of the DOE or the U.S. government.

# Supplemental Material for

# Detection and modeling of hole capture by single point defects under variable electric fields

Artur Lozovoi[1], YunHeng Chen[3], Gyorgy Vizkelethy[4], Edward Bielejec[4], Johannes Flick[1,2,5], Marcus W. Doherty[3], Carlos A. Meriles[1,2,*]

[1]Department. of Physics, CUNY-City College of New York, New York, NY 10031, USA.
[2]CUNY-Graduate Center, New York, NY 10016, USA.
[3]Department of Quantum Science and Technology, Research School of Physics, Australian National University, Canberra, Australian Capital Territory 2601, Australia.
[4]Sandia National Laboratories, Albuquerque, New Mexico 87185, USA
[5]Center for Computational Quantum Physics, Flatiron Institute, New York, NY 10010, USA.
[*]Corresponding author. E-mail: cmeriles@ccny.cuny.edu.

**Experimental setup**

Optical measurements are carried out on a home-built confocal microscope with a free space objective (NA=0.7) equipped with three continuous-wave (cw) diode lasers with emission at 520 nm, 594 nm (Coherent), and 632 nm; in all cases, we reach a diffraction limited illumination spot (with diameter of order 0.5 μm). The laser beams are combined with the use of 605 nm and 550 nm long-pass dichroic mirrors (Semrock) into a single-mode fiber (Thorlabs). The 520-nm laser allows for fast pulsing (100 MHz), whereas the 594-nm beam is controlled via an acousto-optic modulator with <30 ns switching times (Isomet). A single-photon counting module based on an avalanche photodiode (Excelitas) is used for collection. A 650 nm long-pass dichroic mirror (Semrock) separates the excitation light from the detected photoluminescence, which is then coupled into a 9 μm single-mode fiber (Thorlabs). The electric field is applied via a homebuilt switch based on a fast MOSFET (8N120K5), which allows for rise and fall times of ~100 ns for 50 V on and off. All experiments are carried out under ambient conditions.

**NV charge state readout.**

Throughout the present experiments, we implement a single-shot optical readout of the NV charge state. It relies on counting the number of photons detected under a low-power orange 594 nm laser pulse, which weakly excites the NV. The recorded fluorescence can be spectrally filtered to discriminate between NV$^-$ (>637 nm) and NV$^0$ (>575 nm). We use a 650 nm long-pass filter to detect NV$^-$ as the bright state and filter out NV$^0$ as the dark state. The length of the orange laser pulse is chosen based on the following two factors: On the one hand, a longer duration allows for more photoluminescence photons to be detected, while on the other, it causes the charge state switch with a higher probability thereby compromising the discrimination between the two states. In Fig. S1, a photon histogram under 20 ms of 10 μW, 594 nm is presented. Dark gray bars correspond to the detection after high-power 632nm laser pulse that fully ionizes NV into NV$^0$. Blue bars are recorded after a 1mW, 520 nm laser pulse that produces NV$^-$ with ~75% probability. We can clearly see the two overlapping distributions comprising the dark and the bright states. Setting the cut-off to 200 cts/s allows us to assign the charge state in a single shot with >90% fidelity.

**Overview of the theoretical formalism.**

The Monte Carlo simulation of hole transport implemented in this work is based on the approach described by Jacoboni and Reggiani [1]. The key addition is a treatment of a single, positively charged hole in the vicinity of a single negatively charged trap in the absence of other carriers or traps in the crystal, which results in an unscreened long-range Coulomb attraction responsible for the experimentally observed hole capture cross section. The motion of the particle between the scattering events is treated as classical and is described by Newtonian equations of motion in 3D. The initial velocity of the particle is set to its thermal value for a given temperature and the initial direction of motion is chosen randomly with an isotropic distribution in space for each iteration of the simulation ($N = 50000$ iterations for a given set of conditions). Scattering events are introduced into the simulation statistically according to the probability distributions described in detail below.

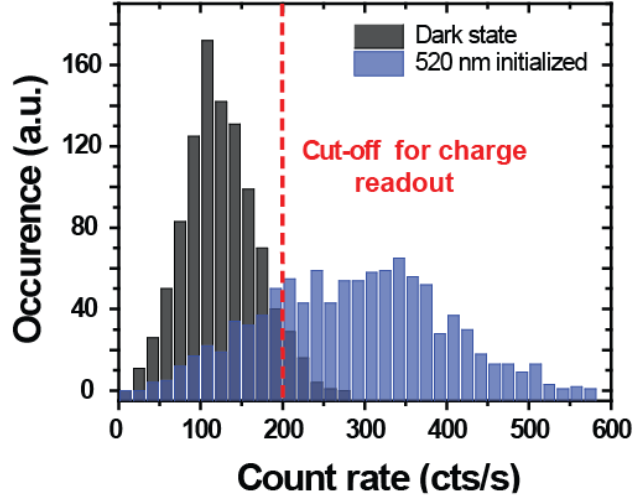

**Figure S1.** Photon histograms acquired under 20 ms/10 μW 594nm illumination of the target NV after 6 mW 632nm (dark gray) and after 1 mW of 520nm illumination (blue). Red dashed line shows the cutoff used for a single-shot charge state discrimination.

*Acoustic phonon scattering.*

The following is the expression for the acoustic phonon scattering probability per unit time for a hole to scatter from the state with a wavevector $k$ to the state $k'$:

$$P_{h,ac}(\boldsymbol{k}, \boldsymbol{k}') = \frac{\pi q (W_1^0)^2}{V \rho u} \binom{N_q}{N_q+1} \frac{1}{4}(1 + 3\cos^2\theta) \delta\left(E_k(\boldsymbol{k}') - E_k(\boldsymbol{k}) \mp \hbar q u\right), \tag{S1}$$

where $\theta$ is the angle between $\boldsymbol{k}$ and $\boldsymbol{k}'$, $N_q = \frac{1}{e^{\frac{\hbar q u}{k_B T}} - 1}$ is the phonon density of states, $q$ is the phonon wavevector modulus, $u = 12 \frac{km}{s}$ is the sound velocity in diamond, $W_1^0 = 12$ eV is the deformation potential in diamond [2], $V$ denotes the volume, $\rho = 3500 \frac{kg}{m^3}$ is the diamond density, $E_k(\boldsymbol{k})$ is the hole kinetic energy. The top term stands for absorption and the bottom term stands for emission. Eq. (S1) is derived using deformation potential theory where scattering is treated as a perturbation and a warped heavy-hole band is assumed [1]. The factor $\frac{1}{4}(1 + 3\cos^2\theta)$ appears due to the overlap integral for the $p$-like hole wavefunctions at the valence band maximum [3,4]. It constitutes the main difference between the scattering rate expression for the holes and for the electrons, where the overlap integral is 1. Considering only the first order term $\frac{\hbar q u}{k_B T}$, the following relation $q = 2^{1/2} k (1 - \cos\theta)^{1/2}$ can be employed and used to integrate Eq. (S1) over $\boldsymbol{k}'$ leading to:

$$P_{h,ac}(\boldsymbol{k}, \boldsymbol{k}') = \frac{\varepsilon_1^{0^2}}{2^{1/2} \pi^2 \rho u} \binom{N_q}{N_q+1} (1 + 3\cos^2\theta) k \frac{(1-\cos\theta)(E_k \pm E_{ph})^{1/2}}{(a(1-g(\vartheta',\psi')))^{3/2}} \tag{S2}$$

Here, $E_{ph} = \hbar qu$ is the phonon energy, $a$ is a non-parabolicity parameter describing the valence band of diamond, $g(\vartheta', \psi')$ is a function containing the angular dependence on the polar $\vartheta'$ and azimuthal $\psi'$ angles, and the valence band parameters [5].

Next, the integration of Eq. (S2) over a solid angle must be carried out to obtain the integrated scattering probability. Here, two important approximations are made [6]. First, the nearly elastic approximation: $(E_k \pm E_{ph})^{1/2} \cong E_k^{1/2}(1 \pm \frac{E_{ph}}{2E_k})$. Second, the small warping approximation: $a(1 - g(\vartheta', \psi')) = \frac{\hbar^2}{2m_h}$ - which substitutes the density-of-states effective mass for a warping-dependent effective mass. Both of these approximations are meant to substitute the standard elastic ($\frac{E_{ph}}{E_k} \ll 1$) and energy equipartition ($\frac{E_k}{k_B T} \ll 1$) approximations that are not valid at low temperatures (T < 77 K), for which the results are presented in Fig. 4(d) in the main text. The direction of $k$ is taken as a polar axis for angular integration. The result of the integration is then as follows:

$$P_{h,ac}(E_k) = \frac{(W_1^0)^2 m_h^{1/2} (k_B T)^3}{2^{9/2} \pi \rho u^4 \hbar^4 E_k^{1/2}} \left( \frac{F_1(x) + (k_B T/E_k) F_2(x)}{G_1(x) - (k_B T/E_k) G_2(x)} \right) \quad \text{(S3)}$$

Here, $m_h$ is an effective hole mass (here taken as 90% of the electron mass as an average between 111, 100 and 110 directions [2] in a single heavy-hole band approximation), $x = (2 m_h^{1/2} u E_k^{1/2})/(k_B T)$ is a dimensionless hole energy, and $F_{1,2}(x)$ and $G_{1,2}(x)$ are expressions dependent on the phonon number distribution [3], which we evaluate numerically in Matlab [5,6]:

$$F_3(x) = \int_0^{\sqrt{2}x} \frac{z^2 \left(1 + 3\left(1 - \frac{z^2}{x^2}\right)^2\right)}{\exp(z) - 1} dz$$

$$F_4(x) = \int_0^{\sqrt{2}x} \frac{z^3 \left(1 + 3\left(1 - \frac{z^2}{x^2}\right)^2\right)}{\exp(z) - 1} dz$$

$$G_3(x) = \int_0^{\sqrt{2}x} \frac{z^2 \left(1 + 3\left(1 - \frac{z^2}{x^2}\right)^2\right)}{1 - \exp(-z)} dz \quad \text{(S4)}$$

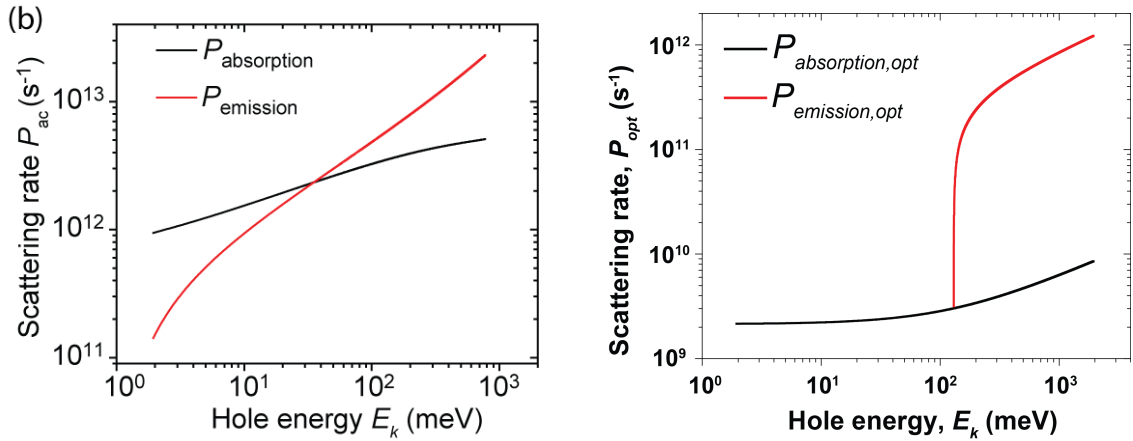

**FIG. S2. Schematics of the Monte Carlo simulation.** (a) Calculated scattering rate for acoustic phonon absorption and emission in room temperature diamond as a function of the hole energy as defined by Eq. (S3). (b) Calculated scattering rate for optical phonon absorption and emission as defined by Eq. (S6). Emission rate drops to zero if the energy of the hole is below the energy of the optical phonon in diamond (175 meV).

$$G_4(x) = \int_0^{\sqrt{2}x} \frac{z^3\left(1+3\left(1-\frac{z^2}{x^2}\right)^2\right)}{1-\exp(-z)} dz$$

The acoustic phonon absorption and emission rates as a function of hole energy at room temperature are shown in Fig. S2(a).

In the Monte Carlo simulation, the intervals between the acoustic scattering events are chosen randomly according to the distribution $P_{h,ac}(E_k)$. When a scattering event takes place, a new $k'$ vector is determined. For this, two random angles $\vartheta'$ and $\psi'$, are generated and used to get an angle between $k$ and $k'$ as $\cos\theta = \cos\vartheta\cos\vartheta' + \sin\vartheta\sin\vartheta'\cos(\psi-\psi')$. With the use of this angle, the phonon wavevector can be calculated as $q = 2^{1/2}k(1-\cos\theta)^{1/2}$. Subsequently, the values of the two angles $\vartheta'$ and $\psi'$ are either accepted or rejected when the value of $q$ derived from them is compared with the acoustic phonon scattering probability given by Eq. S2. This constitutes a Monte Carlo rejection technique that stochastically produces values satisfying a given distribution function depending on the input parameters ($\vartheta'$ and $\psi'$ in our case). As a result of this procedure, the final hole state as well as the phonon energy is determined in accordance with the proposed model. These quantities are finally used as initial conditions for the next free-flight, scattering-free simulation period.

*Optical phonon scattering.*

Optical phonon scattering is taken into account only for the intra-band transitions. Its probability per unit time is:

$$P_{h,op}(k,k') = \frac{3\pi d_0^2}{2\rho V \omega_{op} a_0^2}\binom{N_{op}}{N_{op}+1}\delta\left(E_k(k') - E_k(k) \mp \hbar\omega_{op}\right), \tag{S5}$$

where $d_0 = 4*10^{10}$ eV [7] is the optical deformation potential, $\omega_{op} = 175\ meV$ is the energy of the optical phonon in diamond (Ref. 48 in the main text), $a_0 = 0.357$ nm is the lattice constant of diamond. The top term stands for absorption and the bottom term stands for emission. Integration of Eq. (S5) over $k'$ and then over the solid angle results in:

$$P_{h,op}(E_k) = \frac{3d_0^2 m_h^{3/2}}{2\rho\pi\hbar^3\omega_{op}}\binom{N_{op}}{N_{op}+1}\left(E_k \pm \hbar\omega_{op}\right)^{1/2}, \tag{S6}$$

where $m_h^{3/2} = \frac{m_0^{3/2}}{4\pi|A|^{3/2}}\int_0^{2\pi}d\psi'\int_0^\pi \sin\vartheta'\left(1-g(\vartheta',\psi')\right)^{-3/2}d\vartheta'$ is evaluated numerically in MATLAB with $|A| = 3.61$ being the hole band parameter in diamond [1]. The optical phonon absorption and emission rates as a function of hole energy at room temperature are shown in Fig. S2(b). The emission rate drops to 0 when the hole energy is smaller than the energy of the optical phonon in diamond, 175 meV.

*Externally applied electric field.*

Including the externally applied electric field into the Monte Carlo simulation is done by introducing an electric field force term into the Newtonian equation of motion. The magnitude of this term depends on the voltage applied to the electrodes: $F_{el} = eE = e\frac{V}{d}$, where $d$ is the distance between the electrodes. The direction in which the term acts is constant and parallel to the line connecting the initial position of the hole and the position of the trap.

**Initial distance and energy cutoff in the simulation**

One of the important parameters in the Monte Carlo simulation is the initial distance between the hole and the trap. In the experiments presented in the main text, the distance between the source and the target NVs is 3.9 μm. Transport simulation in the volume of the corresponding size is computationally too demanding. Therefore, we run the simulation of a hole capture probability as a function of electric field strength for a series of increasing initial distances in order to determine whether the results converge to the same form after certain separation. They are shown in Fig. S3. The presented dependence stops changing for initial distances above 350 nm, thus justifying the use of 450 nm as the initial distance between the hole and the NV for the simulations presented in the main text allowing extrapolation of the results to the conditions of the experiments. At temperatures below 60 K, the initial distance at which the results converge is larger due to a drastic increase in the capture cross section values ($> 300 \times 300$ nm$^2$). At 10 K, the appropriate initial distance is estimated to be 3000 nm as demonstrated by the convergence of the simulation results in Fig. S3 (b). In Fig. 4(d) of the main text, we use this initial distance for the simulations in the temperature range $10 \div 60$ K.

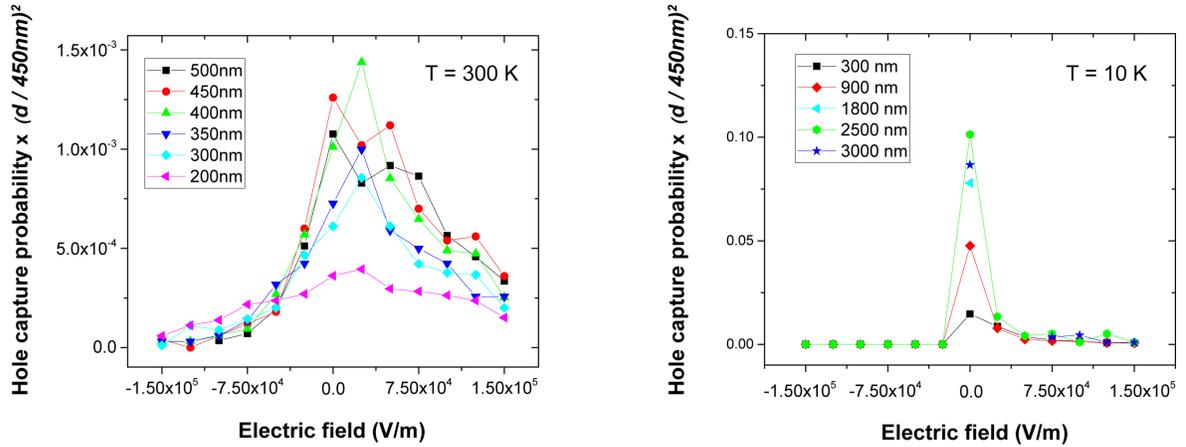

**Figure S3.** Calculated capture probabilities as a function of the electric field strength at room temperature (a) and 10 K as extracted from the Monte Carlo simulation for different initial hole–trap distances.

The energy cutoff employed to consider the hole as captured (See Fig. 4a of the main text) is a rather arbitrary parameter, the choice of which should be tentatively related to the energy of the optical phonon in diamond (175 meV). This consideration stems from the requirement that the scattering event involving an optical phonon absorption by the hole should not transfer enough kinetic energy to the carrier so that it can leave the bound energy states. If this condition is fulfilled, then it is guaranteed that the hole gets captured by the trap upon reaching the negative total energy below the cutoff value. Based on this, we expect that further increasing the cutoff will not change the capture cross section extracted from the simulation, which is taken as a sign that the optimal cutoff values have been reached. In Fig. S4, this exact scenario is observed in the dependence of the capture cross section on the value of the energy cutoff calculated at room T. The onset of the plateau is observed around 100 meV - below the optical phonon energy. Here, we ignore the details of how exactly the hole goes down the energy level ladder below the cutoff to eventually be captured by NV$^-$ and how much time it spends at each of the levels. We also consider the shift of the bound hole energy levels due to the presence external field to be negligible for the field strengths used in the current work. Both the problems were treated in [8].

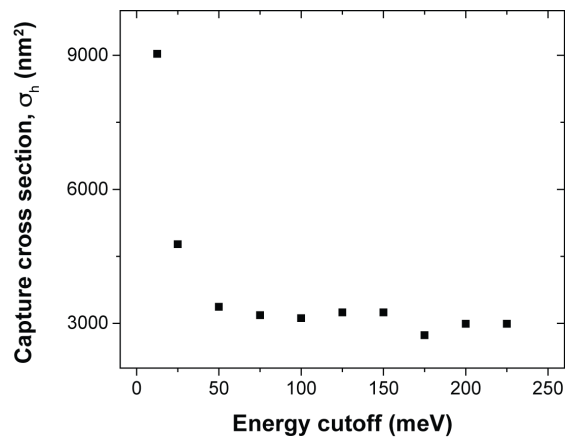

**Figure S4.** Hole capture cross section as a function of the energy cutoff for the capture event at room temperature as extracted from the Monte Carlo simulation.

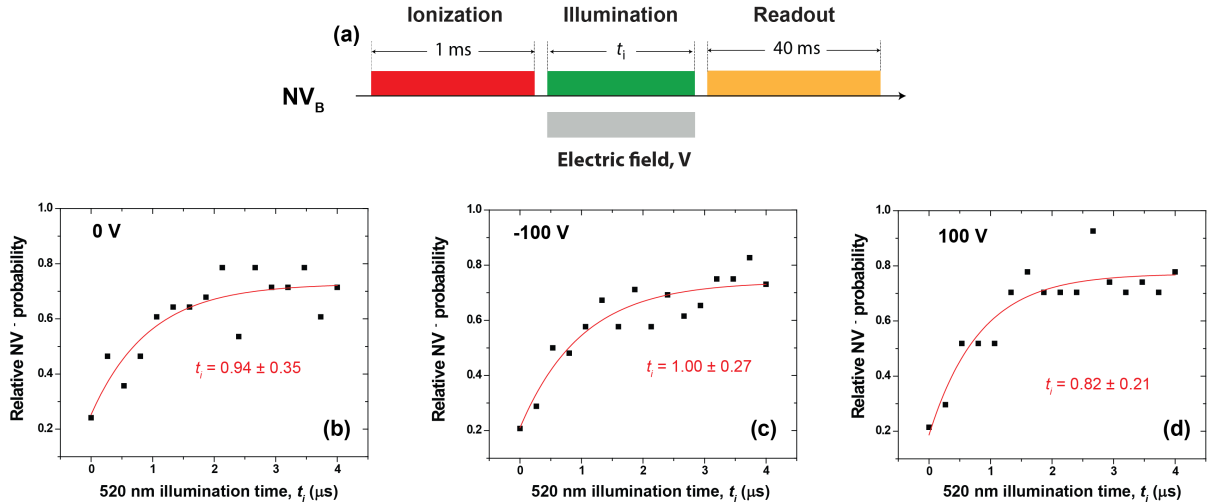

**Figure S5.** (a) Laser pulse sequence used for the measurement of the charge cycling rate of the NV$_B$. Red 632nm (1 ms and 5 mW) is used to prepare it into NV$^0$; 520 nm of variable duration (1 mW) recombines NV$^0$ into predominantly (~ 75%) NV$^-$ with or without the application of the electric field; 594 nm illumination (40 ms and 12 µW) is used to read out the charge state. (b), (c) and (d) show recombination dynamics under 1 mW of 520 nm illumination in the absence of the electric field (b) and in the presence of -100 V (c) and 100 V (d). All three curves are fitted with an exponential function yielding the same value of characteristic charge state cycling time ~0.9 µs.

**NV ionization and recombination in the presence of the electric field.**

Calculation of the hole capture cross section based on the transport experiment presented in the main text relies on the knowledge of the ionization and recombination rates of NV$_B$ that acts as a source of the free charge carriers. The experimental protocol used to measure these rates is shown in Fig. S5 (a). After ionizing the NV into the neutral charge state with high-power 632 nm illumination, we extract the characteristic time constant of recombination under 1 mW of green 520 nm illumination in the absence of the electric field (Fig. S5 (b)), in the presence of the decelerating -100 V (Fig. S5 (c)) and in the presence of accelerating 100 V (Fig. S5 (d)) corresponding to the maximal electric field strength applied in the transport experiment. Here, a single-shot charge state readout under 40 ms of 594 nm (12 µW) illumination is employed to measure the probability to find the defect in NV$^-$. Lower fidelity of the threshold between the dark and the bright state of NV$_B$ compared to NV$_A$ is responsible for a non-zero NV$^-$ population after the initial red ionization pulse. In all three cases (Fig. S5 (b), (c) and (d)), the characteristic recombination time is found to be ~0.9 µs within the error margin. We note that in the transport experiments under the AC field reported in Fig. 2(b) of the main text the time period during which the field is on is 1 µs - on the order of the ionization and recombination time of the NV$_B$. Therefore, the electric field can be treated as DC during each period of the NV charge state interconversion. Based on this, we conclude that in the current sample the charge state dynamics of the NV is unaffected by the application of the external electric fields within the range of field strengths used in the transport experiments.

**The effect of the sample holder on the shape of the capture probability distribution.**

In the experiments presented in the main text and in the Supplemental Materials up until this point, the diamond sample was mounted on a grounded copper mount, which inevitably influences the value and direction of the external electric field applied on the electrodes. This effect is further exacerbated by the large size of the electrodes (1×2 mm) used in these experiments and the big distance between them (1 mm). In Fig. 6 we present a set of measurements carried out on the same pair of NVs separated by 3.9 μm that was examined in Figs. 1 and 2 of the main text; however, the diamond here was mounted on a PCB-board over an omega shaped microwave waveguide with a diameter of ~1.5 mm. The two main features of the dependence remained the same, namely, the peak at 0 V and the overall asymmetry though the tail in the positive (accelerating) electric field range is noticeably longer compared to the dependence shown in the main text. We tentatively attribute this dissimilarity to changes in the electric field value and direction due to screening caused by the sample mount, different in the two sets of the experiments. The ideal scenario would require significantly smaller electrodes fabricated locally on the surface of the diamond next to the NVs of interest. This would allow to mitigate the effect of conducting materials around the sample on the geometry and strength of the applied electric field, and possibly minimize the build-up of space-charge potentials.

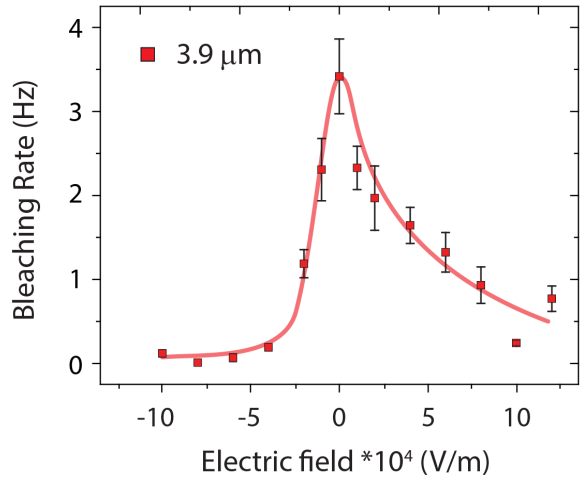

**Figure S6.** Hole capture unit-time probability as a function of the applied AC voltage for the same NV pair shown in Figs. 1 and 2 of the main text but mounted on a PCB-board with an omega-shaped microwave waveguide underneath it. The asymmetry of the distribution is more pronounced compared to the data shown in Fig. 3 of the main text. Solid line is a guide to the eye.